\documentclass[english,aps,prl,twocolumn,superscriptaddress,showpacs]{revtex4}
\usepackage{graphicx}
\usepackage{babel}
\usepackage{epstopdf}

\begin{document}

\title{Nonaxisymmetric Anisotropy of Solar Wind Turbulence as a Direct Test  for  Models of Magnetohydrodynamic Turbulence}

\author{A.J.~Turner}
\email{a.j.turner@warwick.ac.uk}
\affiliation{Centre for Fusion, Space and Astrophysics; University of Warwick,
Coventry, CV4 7AL, United Kingdom}

\author{G.~Gogoberidze}
\affiliation{Centre for Fusion, Space and Astrophysics; University of Warwick,
Coventry, CV4 7AL, United Kingdom}
\affiliation{Institute of Theoretical Physics, Ilia State University, 3/5 Cholokashvili ave., 0162 Tbilisi, Georgia}

\author{S.C.~Chapman}
\affiliation{Centre for Fusion, Space and Astrophysics; University of Warwick,
Coventry, CV4 7AL, United Kingdom}

\begin{abstract}
Single point spacecraft observations of the turbulent solar wind flow exhibit a characteristic nonaxisymmetric anisotropy that depends sensitively on the perpendicular power spectral exponent. For the first time we use this nonaxisymmetric anisotropy as a function of wave vector direction to test models of magnetohydrodynamic (MHD) turbulence.
Using Ulysses magnetic field observations in the fast, quiet polar solar wind
we find that the Goldreich-Sridhar model of MHD turbulence is not consistent with the observed anisotropy, whereas the observations are well reproduced by the ``Slab + 2D" model. The  Golderich-Sridhar model  alone can not account for the observations unless an additional component is also present.

\end{abstract}


\maketitle

In-situ satellite observations of the solar wind show a broad spatio-temporal range of plasma fluctuations \cite{Coleman_1968} and a high effective magnetic Reynolds number \cite{Matthaeus_2005, Weygand_2009}. This suggests a nonlinear cascade consistent with turbulence \cite{Tu_et_al_1984, Tu_Marsch_1995}. In-situ observations have hence been used extensively to test the predictions of Magnetohydrodynamic (MHD) turbulence \cite{Matthaeus_1995, Bieber_1996, Bruno_2005, Chandran_2000, WEa11, Forman_2011}. In the solar wind the background magnetic field imposes a preferred direction in the plasma resulting in an inherent anisotropy in the turbulent fluctuations \cite{Belcher_1971, Bavassana_1982}. This anisotropy is fundamental to the understanding of cosmic ray scattering \cite{Chandran_2000} and the non-adiabatic rate of cooling observed for the solar wind \cite{SEa09}.

A pioneering anisotropic model of MHD turbulence is that of Goldreich and Sridhar \cite{GS95} (GS model hereafter). In the GS model, the turbulence is dominated by the cascade perpendicular to the local mean magnetic field which follows Kolomogorov phenomenology with a one-dimensional perpendicular energy spectrum  predicted to be $E(k_\perp)\sim k_\perp^{-5/3}$.
 A ``critical balance"  between the linear and nonlinear time-scales  results in  a one-dimensional parallel spectrum of $E(k_\parallel)\sim k_\parallel^{-2}$ for the GS model. Here $k_\perp$ and $k_\parallel$ are the components of wave vector ${\bf k}$ perpendicular and parallel to the local magnetic field direction. The energy spectrum is intrinsically axisymmetric w.r.t. the local magnetic field.
 There is significant numerical and observational support for the predictions made by the GS model.
 This support includes high resolution direct numerical simulations (DNS) \cite{MG01} and recent solar wind observations that confirm predictions of the anisotropy of the spectral exponents \cite{Horbury_2008,Podesta_2009,WEa11}.
 Paradoxically, the GS model is inefficient in scattering cosmic rays due to the inherent strong anisotropy ($k_\perp \gg k_\parallel$) at small scales, which cannot account for the observed scattering characteristics \cite{Chandran_2000}.

  A possible resolution of this inconsistency is that  there is an additional `slab' component of solar wind fluctuations  that is independent of the perpendicular cascade, consisting  of fluctuations with wave vectors parallel to the local magnetic field. There is considerable observational evidence that in the solar wind turbulence there is a combination of perpendicular and parallel wave numbers. Classically this is seen in  the magnetic field correlation function which, assuming axisymmetry, gives a  `Maltese cross'  pattern \cite{Matthaeus_1990,Matthaeus_1995}. This result was interpreted as a superposition of two-dimensional (fluctuations with ${\bf k}$ purely perpendicular to the local magnetic field) and slab components \cite{Matthaeus_1990} (S2D model hereafter). The observed anisotropy of the spectral exponents that support the GS model in the solar wind do not exclude  the S2D model \cite{Horbury_2008,Podesta_2009,WEa11}. It is thus an open question as to whether the GS model operates as the sole turbulence process in the solar wind.

Belcher and Davis \cite{Belcher_1971} used Mariner 5 observations to investigate the anisotropy of solar wind magnetic fluctuations in the low frequency (energy containing) and inertial intervals. They observed that the power in the fluctuations is ordered \emph{both} w.r.t. the average magnetic field \emph{and} the solar wind flow direction. If the solar wind turbulence is described by the S2D model with the same spectral exponents for both components, this observed nonaxisymmetric anisotropy can be explained as a sampling effect \cite{Bieber_1996} related to the Taylor hypothesis \cite{Taylor}. Importantly, if the parallel and perpendicular components have distinct power spectral exponents, as in models such as GS, then the Taylor hypothesis can quite generally relate the exponents to observed nonaxisymmetry \cite{Turner_2011}. 
This suggests a new quantitative test for theoretical predictions,  since the observed nonaxisymmetry does not depend solely upon the power spectral exponents of $E(k_\perp)$ or $E(k_\parallel)$, but as we shall see, is sensitively dependent upon the transition from $k_\perp$ to $k_\parallel$ described by the models. 
In this Letter we will use this idea to test the predictions of the GS model and the S2D model against solar wind observations. We find that the S2D model can fit the data whereas the GS model cannot.
In addition to critically balanced turbulence some other component must be  present in the solar wind fluctuations to account for the observations.

Theories of turbulence predict how the energy in the fluctuations vary with scale. For fluctuations in a vector field, the  Fourier transform of the two point correlation matrix $R_{ij}({\bf r})=\langle \delta B_i({\bf x}) \delta B_i({\bf x}+{\bf r})\rangle$ then defines a spectral energy density \emph{ tensor} $P_{ij}(\mathbf{k})$ \cite{Turner_2011, Forman_2011} which captures the full anisotropy of the fluctuations.  Single point observations in the flow cannot isolate ${\bf k}$ uniquely and instead give a reduced one-dimensional spectral tensor $\tilde{P}_{ij}(f)$. From Taylor's hypothesis \cite{Taylor}, the measured one-dimensional tensor $\tilde{P}_{ij}(f)$ is the integral of $P_{ij}(\mathbf{k})$   over the plane $\mathbf{k} \cdot \mathbf{V}_{sw}=2\pi f$ in ${\bf k}$ space, where $\mathbf{V}_{sw}$ is the solar wind velocity, i.e. we observe:
\begin{equation}
\tilde{P}_{ij}(f,\theta) = \int d^3k P_{ij}(\mathbf{k}) \delta (2\pi f - \mathbf{k} \cdot \mathbf{V}_{sw}). \label{eq:Taylor}
\end{equation}

The observed spectral tensor $\tilde{P}_{ij}(f,\theta)$ is in general nonaxisymmetric, with dependence on both $f$ and the angle $\theta$ between the local magnetic field and the solar wind flow velocity. This is the case even if the underlying turbulence phenomenology is axially symmetric w.r.t. the local magnetic field direction, as is the case for both the GS and S2D models. A natural co-ordinate system \cite{Bieber_1996, Turner_2011, Forman_2011} is to project the magnetic fluctuations onto a local, scale  dependent mean field direction
$\mathbf{e}_{z}(t,f)=\overline{\mathbf{B}}(t,f)/\left|\overline{\mathbf{B}}(t,f)\right|$ where $\overline{\mathbf{B}}(t,f)$ is the scale dependent local average field. If the
bulk flow velocity unit vector is $\hat{\mathbf{V}}$ then
the other two perpendicular unit vectors of the set are:
\begin{equation}
\mathbf{e}_{x}(t,f)= \frac{ \mathbf{e}_{z}\times \hat{\mathbf{V}}}{ \left|\mathbf{e}_{z}\times \hat{\mathbf{V}} \right|},~~
\mathbf{e}_{y}(t,f)=\mathbf{e}_{z}\times\mathbf{e}_{x}. \label{eq:e_BV}
\end{equation}
We will focus on the power ratio in this coordinate system:
\begin{equation}
R(\theta,f) = \frac{\tilde{P}_{xx}(\theta,f)}{\tilde{P}_{yy}(\theta,f)}. \label{eq:Ratio}
\end{equation}
 We now calculate (\ref{eq:Ratio}) for the GS and S2D models and compare with observations.

\emph{GS model:} We will assume that for the GS model all parallel components of the spectral tensor vanish so that $P_{z i} \equiv 0$, consistent with other studies \cite{Bieber_1996}.
The perpendicular components for the GS model are related to the power tensor in the following manner:
\begin{equation}
P_{ij}^{GS}({\bf k})=\frac{E(\mathbf{k}) }{4 \pi k_\perp^2} \Pi_{ij}, \label{eq:CB_PSD}
\end{equation}
where $\Pi_{ij} = \delta_{ij} - {k}_{\perp i}{k}_{\perp j}/k^2_\perp$
 and from Ref. \cite{GS95}:
\begin{equation}
E(\mathbf{k}) = C_K \frac{\varepsilon^{2/3} L^{1/3} }{k_\perp^{10/3}} \emph{g}\left( \frac{k_\parallel L^{1/3}}{k_\perp^{2/3}} \right). \label{eq:CB}
\end{equation}
Here $C_K$ is the Kolmogorov constant, $L $ is the characteristic injection scale, $\varepsilon$ is the energy dissipation rate and $g$ is a positive symmetric function related to the scaling between $k_\parallel$ and
 $k_\perp$, where $g(0) = 1$ and $\int_0^\infty g(z)dz = 1$ \cite{GS95}. The theoretical prediction of the GS model is determined by substitution of Eqs. (\ref{eq:Taylor}), (\ref{eq:CB_PSD}) and (\ref{eq:CB}) into Eq. (\ref{eq:Ratio}) and numerical integration. We find that the result is not sensitive to the functional form of the scaling function $g$ and we use the exponential function
[$g=\exp(-L^{1/3}|k_\parallel|/k_\perp^{2/3})$].

The GS model, as discussed here, assumes balanced MHD turbulence with equal power in the  Alfv\'en waves travelling in both directions along the local magnetic field. Turbulence in the fast solar wind is known to be imbalanced with more power in the Alfv\'en waves propagating outward from the sun than toward it \cite{Bruno_2005}. However, an imbalanced extension \cite{Lithwick_2007} predicts the same spectral exponents of both dominant and subdominant waves as well as the same scaling relation between $k_\parallel$ and $k_\perp$ as the balanced GS model. Thus, given isotropy of the turbulence at the injection scale $L$, imbalance of the turbulence does not affect the ratio (\ref{eq:Ratio}).

\emph{S2D model:} We follow the prescription of  Ref. \cite{Bieber_1996}, except that we specify distinct spectral exponents $q_s$ and $q_{2D}$ for the slab and 2D components as is observed. In this Letter, we will consider variation in $q_{2D}$, as $R(\theta,f)$ is sensitive to the perpendicular spectral exponent. For simplicity a constant value of $q_s=2$ is used throughout \cite{WEa11, Turner_2011}. In this case the ratio $R_{S2}(\theta,f)$ predicted by the S2D model is
\begin{eqnarray}
R_{S2}(\theta,f) = \frac{A_c \left( \frac{2\pi f L}{V_{sw} \cos \theta} \right)^{1-q_s} + q_{2D} \left( \frac{2\pi f L}{V_{sw} \sin \theta} \right)^{1-q_{2D}}}{A_c \left( \frac{2\pi f L}{V_{sw} \cos \theta} \right)^{1-q_s} + \left( \frac{2\pi f L}{V_{sw} \sin \theta} \right)^{1-q_{2D}}}, \label{eq:S2_ratio}
\end{eqnarray}
where $L$ is the injection scale (we assume that both components have the same injection scale) and $A_c$ is a constant that characterizes the energy ratio of slab and 2D components. Since $q_s \ne q_{2D}$ the relative power of the components is $f$ dependent, as is the ratio $R_{S2}(\theta,f)$.
Observations  \cite{Bieber_1996} give a $1:4$  energy ratio between slab and 2D components and we will fix this at the injection scale $L$, which fixes the constant $A_c \approx 0.5$.

\begin{figure*}
\begin{centering}
\includegraphics[width=2\columnwidth]{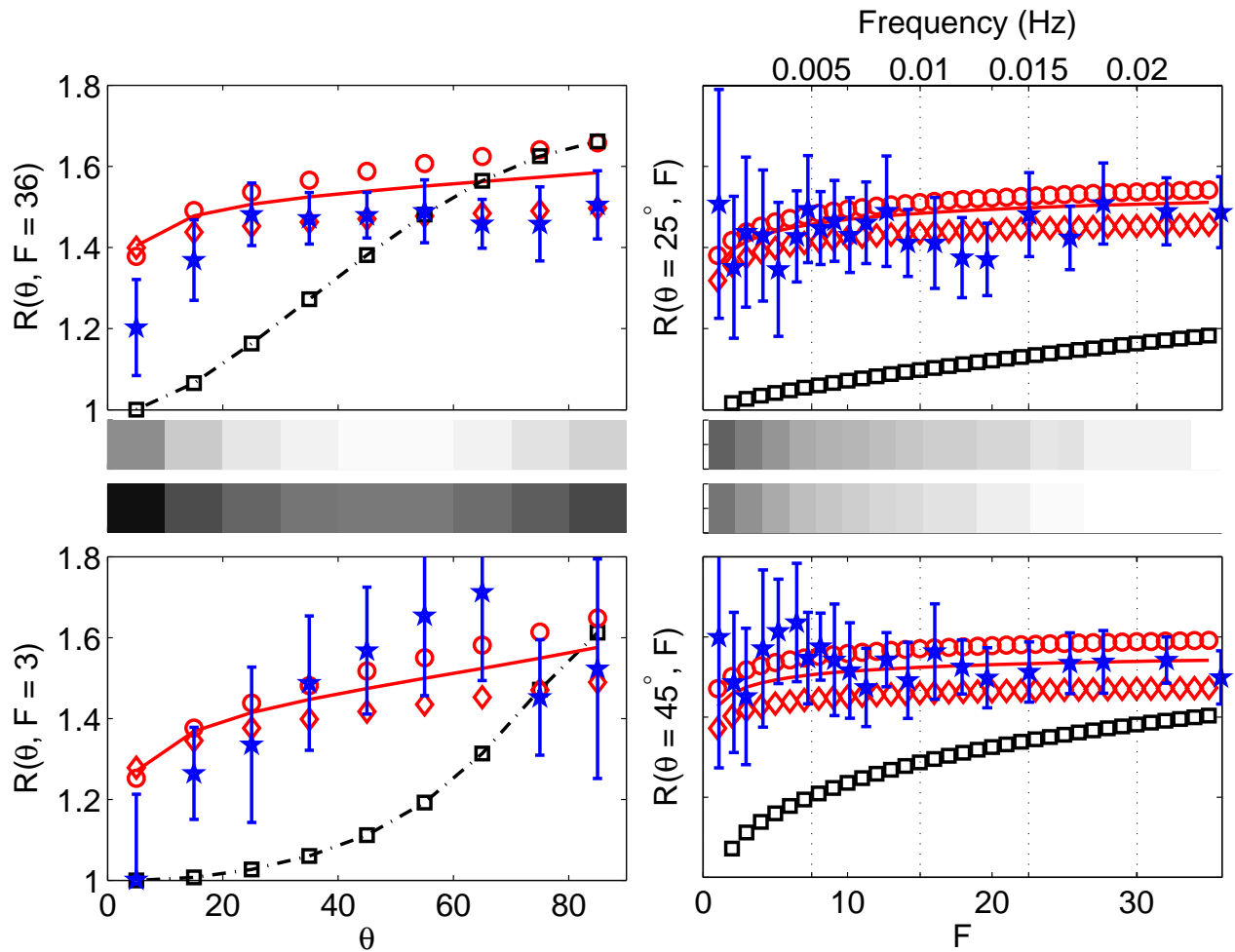}
\par\end{centering}
\caption{Left panels - $\theta$ dependence of the ratio $R(\theta,F)$ at two normalised frequencies: $F=36$ (upper plot) and $F=3$ (lower plot). Right panels -- frequency dependence of the ratio $R(\theta,F)$ at $\theta=25^\circ$ (upper plot) and $\theta=45^\circ$ (lower plot). Observations are shown by blue stars, S2D model is in red with circles for Kolmogorov and diamonds for Iroshnikov-Kraichnan perpendicular scaling. A solid red line indicates the S2D model with the  perpendicular scaling exponent from the data of $q_{2D} = 1.59$. All S2D models shown here use $q_s = 2$. The GS model is shown by black rectangles. Middle panels - Number of samples to form $\overline{R}$, where white $= 1800$ and black $= 36$ - this varies significantly with both frequency and angle.  } \label{fig:freq}
\end{figure*}

We plot these theoretical predictions in Figure 1, for $\theta = [5^\circ, ~85^\circ]$, as co-ordinate system (\ref{eq:e_BV}) becomes undefined as $\theta \rightarrow 0^\circ$. The ratio $R$ is plotted using normalized frequency $F=2\pi f L/V_{sw}$ and we show
two different cuts through the surface $R(\theta,F)$; for $F=36$ (upper plot) and $F=3$ (lower plot) in the left panels and for $\theta=25^\circ$ (upper plot) and $\theta=45^\circ$ (lower plot) in the right panels.  The S2D model is indicated by red symbols on the plot,  with circles for Kolmogorov $q_{2D}=5/3$ and diamonds for Iroshnikov-Kraichnan $q_{2D}=3/2$. The GS model prediction is indicated by the black squares.

For underlying turbulence that is axisymmetric, with  $E(k_\perp)\sim k_\perp^{-\gamma_\perp}$ 
the observed $R(\theta,f)\rightarrow \gamma_\perp$ for $\theta\rightarrow 90^\circ$ for all of these curves \cite{Turner_2011}.  
It can be seen in the left panels of Figure \ref{fig:freq} that the predicted $R(\theta,f)$ for the GS and S2D models are distinct for intermediate values of $\theta$ in the transition $R(\theta = 0^\circ \rightarrow 90^\circ,f)$. The GS model form is concave, whereas the S2D model is strongly convex. This provides a strong test against observations provided the statistical variability is smaller than the difference between the predicted curves. Importantly, the model predictions are maximally distinct for intermediate angles (i.e. $\theta \sim 20^\circ-40^\circ$) and as we shall see, this is where the observations tend to be more statistically significant as there more samples.

\emph{Observations:} We use magnetic field data at 1 second resolution for an interval [day 91-146, 1995] of fast solar wind observed by Ulysses. In this interval Ulysses moved from a heliographic latitude of $21^\circ$ to $58^\circ$ and a radial distance of 1.36 to 1.58 AU. This long quiet interval of 55 days is needed to obtain good statistical coverage across $\theta$.
This interval is of fast solar wind with an average flow speed of $756$ km/s and average plasma parameters: magnetic field $\left|\overline{\mathbf{B}}\right| \simeq 2.9$ nT,
ion plasma $\beta \simeq 1.35$,
ion plasma density $n_i \simeq 1.23$ cm$^{-3}$,
ion temperature $T_i \simeq 21$ eV
and Alfv\'en speed $\simeq 56$ km/s.

We use the continuous wavelet transform (CWT) with a Morlet wavelet to resolve vector fluctuations $\delta \mathbf{B}(t,f)$ in time $t$ and frequency $f$. The scale dependent local field
$\overline{\mathbf{B}}(t,f)$  is calculated via the convolution of a Gaussian window as  outlined in Ref. \cite{Turner_2011} to obtain the
 unit vector direction
$\mathbf{e}_{z}(t,f)=\overline{\mathbf{B}}(t,f)/\left|\overline{\mathbf{B}}(t,f)\right|$.
 In practice, the polar wind seen by Ulysses is stable and within $3^\circ$ of the radial direction over the entire interval under study, so we replace $\hat{\mathbf{V}}$ by the radial  unit vector $\mathbf{e}_{R}$ here, which corresponds exactly with the coordinate system used by Ref. \cite{Belcher_1971}.
The magnetic fluctuations are then projected to basis (\ref{eq:e_BV}) to determine the power for different components of $\tilde{P}_{ii}(t,f)$ as a function of frequency $f$, such that $\tilde{P}_{ii}(t,f)=2\Delta \delta B_{i}(t,f)^2$, where $\Delta$ is sampling time of the data. 

For comparison between the models and the observations the data is normalized to $F=1$ at the beginning of the inertial range. This corresponds to a time-scale of 25 minutes. Ratio (\ref{eq:Ratio}) and a measure of the statistical variability for four specific cases of $\theta$ and $F$ are indicated in Figure \ref{fig:freq} by the blue stars. The method for estimating the statistical variability in these observations is given below. Given this variability, we can immediately see that the observations correspond quite well to the S2D model whereas the
  GS model fails to predict the observations. Thus, the GS model cannot reproduce the observed inertial range fluctuations in the fast solar wind in the absence of some additional, scaling component of fluctuations. For completeness, we use the data to obtain $q_{2D} = 1.59$, this gives an S2D model result shown by the solid red line.
 
\begin{figure}
\begin{centering}
\includegraphics[width=\columnwidth]{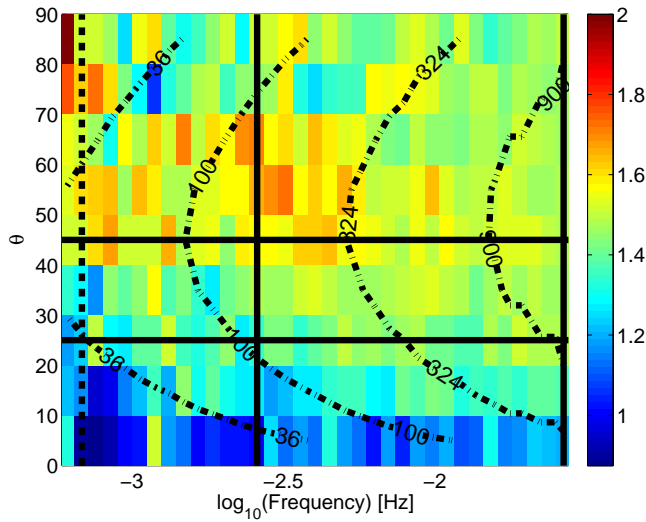}
\par\end{centering}
\caption{ The surface $R( \theta, f)$ is shown in colour. The black solid lines indicate the cuts of $R( \theta, f)$ shown in Figure \ref{fig:freq}. The dashed line shows the start of the inertial range, F = 1. The dash-dot lines show contours of subinterval sample size. The colour bar shows the ratio $R(\theta,f) = \tilde{P}_{xx}(\theta,f)/\tilde{P}_{xx}(\theta,f)$ } \label{fig:R}
\end{figure}
  
The full surface $R(\theta,f)$ for our interval is shown in Figure \ref{fig:R}. This shows how the nonaxisymmetry between the perpendicular directions depends on the sampling domain. The surface is constructed by subdividing the entire interval into 1 day subintervals. $\tilde{P}_{xx}(t,f)/\tilde{P}_{yy}(t,f)$ is binned according to $\theta$ for each of these subinterval. As the distribution in each bin has a form close to log-normal the best measure of the subinterval average, $\overline{R}$, is the geometric mean \cite{Bieber_1996}.
In Figure \ref{fig:R} the contours indicate the number of samples in each $\overline{R}$, thus the contours may be interpreted as confidence contours. Each bin of the surface $R(\theta,f)$ has 55 realisations of $\overline{R}$. These realisations are used to calculate the statistical variability of the distribution $R(\theta,f)$ by calculating the median and interquartile range of the 55 values of $\overline{R}$. Figure \ref{fig:R} shows the median value for each bin and the stars in Figure \ref{fig:freq} show the median value with statistical variability estimated to 99\% certainty indicated by the errorbars.
Observationally, since $\cos\theta = \mathbf{e}_{z} \cdot \mathbf{e}_{R} = \overline{\mathbf{B}}(t,f)/\left|\overline{\mathbf{B}}(t,f)\right| \cdot \mathbf{e}_{R}$ any uncertainty in $\overline{\mathbf{B}}(t,f)$ is greatest as $\theta \rightarrow 0^\circ$. We estimate that an uncertainty of $1\%$ on $\overline{\mathbf{B}}(t,f)$ translates to $\delta \theta \sim 8^\circ$ in the $\theta = 0^\circ - 10^\circ$ interval. Thus the observed values at small $\theta$ are unreliable.

Much of the observational support for GS has relied upon from the determination of spectral exponents. Whilst this is not an unreasonable test for the theories and models of plasma turbulence it is not  complete and may lack uniqueness. Our work highlights the need for  other distinct methods  to test these predictions. Such tests need to probe the full three-dimensional energy spectrum  as the kinematic power ratio used in this Letter begins to do.

\begin{acknowledgments}
The authors acknowledge the Ulysses instrument teams for providing magnetometer data. This work was supported by the UK STFC.
\end{acknowledgments}

\end{document}